\documentclass[aps,prl,showpacs,twocolumn,floatfix,superscriptaddress]{revtex4} 
\usepackage{graphicx}

\begin{document}

\title{Competing ferromagnetic and nematic alignment in self-propelled polar particles}

\author{Sandrine Ngo} 
\affiliation{Service de Physique de l'\'Etat Condens\'e, CEA--Saclay,~91191~Gif-sur-Yvette,~France} 
\affiliation{Max-Planck-Institute for the Physics of Complex Systems, N\"othnitzer Stra{\ss}e 38, D-01187 Dresden, Germany}

\author{Francesco Ginelli} 
\affiliation{Istituto dei Sistemi Complessi, CNR, via dei Taurini 19, I-00185 Roma, Italy}
\affiliation{SUPA, Institute for Complex Systems and Mathematical Biology, King's College, University of Aberdeen, Aberdeen AB24 3UE, United Kingdom}

\author{Hugues Chat\'e} 
\affiliation{Service de Physique de l'\'Etat Condens\'e, CEA--Saclay,~91191~Gif-sur-Yvette,~France} 
\affiliation{Max-Planck-Institute for the Physics of Complex Systems, N\"othnitzer Stra{\ss}e 38, D-01187 Dresden, Germany}

\date{\today} 

\begin{abstract}
We study a Vicsek-style model of self-propelled particles
 where ferromagnetic and nematic alignment compete in both the usual ``metric'' version and in the
``metric-free'' case where a particle interacts with its Voronoi neighbors.
We show that the phase diagram of this out-of-equilibrium XY model is similar to that of 
its equilibrium counterpart: the properties of the fully-nematic model, studied before in
\cite{Ginelli2010}, are thus
robust to the introduction of a modest bias of interactions towards ferromagnetic
alignment. The direct transitions between polar and nematic ordered phases are shown to be 
discontinuous in the metric case, and continuous, belonging to the Ising universality class, 
in the metric-free version.
\end{abstract} 

\maketitle 

Collective motion is a topic currently enjoying interest in various communities
\cite{Sumpter,Giardina-review,SR-review,ABP-review}. 
Within (statistical) physics, the seminal work of Vicsek {\it et al.} \cite{Vicsek1995}, 
followed by the remarkable calculation of Toner and Tu \cite{Toner}, 
has offered to view the emergence of collective motion in leaderless groups of identical individuals
as the spontaneous breaking of rotational invariance. 
The celebrated Vicsek model, which consists of self-propelled particles aligning 
ferromagnetically their orientations with that of their neighbors in the presence of noise, was originally presented ---and rightly so--- as an out-of-equilibrium XY model where spins are forced to move. As is now well-known, 
the Vicsek model is endowed with properties very different from those of the
XY model: in two dimensions, true long-range polar (orientational) order emerges  \cite{Toner}
from a discontinuous phase transition \cite{Chate2008}, and the
long-range correlations and anomalous fluctuations predicted by Toner and Tu for the ordered collective motion phase, although not observed in the region near onset, are indeed present in a large portion of parameter space.

Other, Vicsek-style, flocking models have been introduced which serve as key members of 
different universality classes
for ``dry active matter'' \cite{SR-review}, i.e. situations in which global momentum 
is not conserved and hydrodynamic interactions play no significant role. 
A prominent case is the ``self-propelled rods" model, in which the ferromagnetic
interaction of the Vicsek model is replaced by nematic alignment \cite{Ginelli2010}, 
in line with the typical outcome of inelastic collisions between
moving elongated objects. Switching from ferromagnetic to nematic symmetry of interactions 
in this other out-of-equilibrium XY model, changes the symmetry of the ordered phase 
(which is then nematic), in line with the symmetry change of the quasi-ordered phase of 
the corresponding equilibrium version \cite{GXY}. 
Numerical results \cite{Ginelli2010} suggest that like the original (ferromagnetic) Vicsek model, the nematic order
observed is truly long-range, but no Toner-Tu-like calculation is available to confirm this at some
analytical level.

\begin{figure}
\includegraphics[draft=false,clip=true,width=\columnwidth]{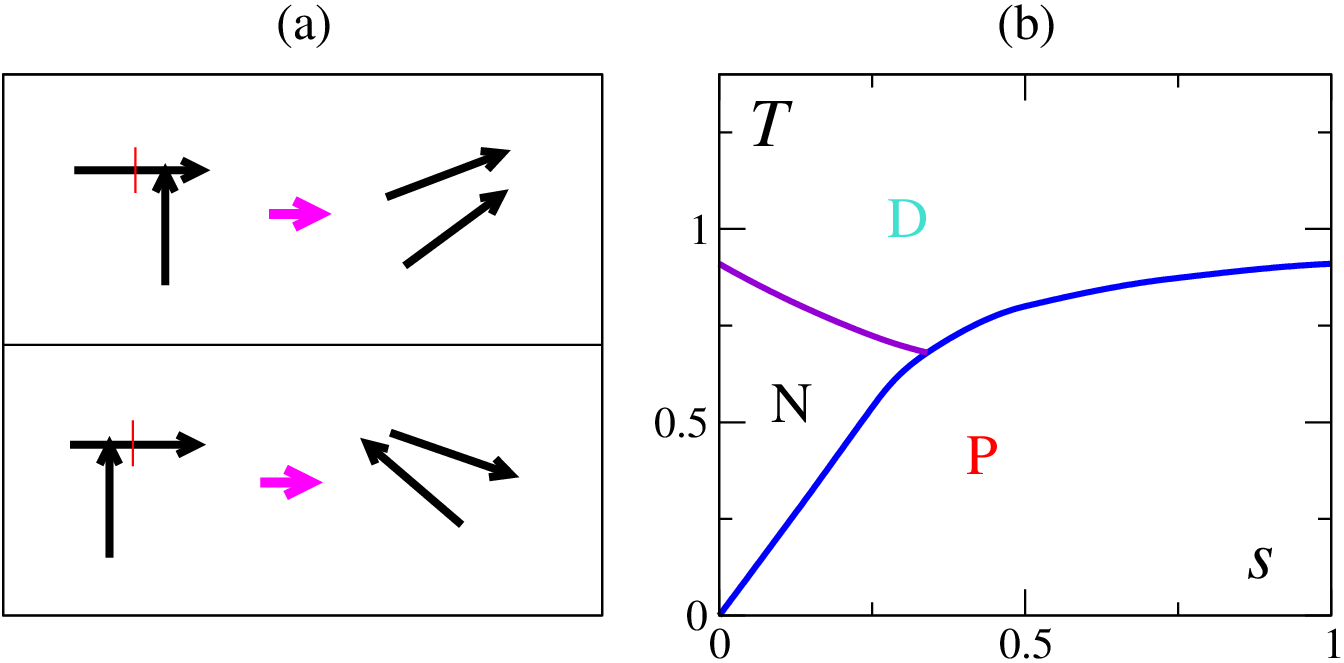}
\caption{(color online) 
(a) Sketch of two self-propelled needle-like objects (black arrows)
moving at the same speed in some overdamped dynamics,  which collide 
with an incoming angle of exactly $\frac{\pi}{2}$. The middle of each rod is indicated by the thin red line.
If the impact point is along the first half of the hit needle, 
(some degree of) polar alignment is expected (top panel), whereas anti-alignment will typically occur
for an impact point at the rear (bottom panel). 
Friction forces may very well, though, lead to polar alignment, not anti-alignment, even if the impact point is slightly beyond the first half of the needle. In such a case,
the nematic symmetry of interactions would be weakly broken and biased towards ferromagnetic
alignment.
(b) Schematic phase diagram of an equilibrium generalized XY model with Hamiltonian
$H=-\sum_{\langle ij\rangle} s\cos[\theta_i-\theta_j] +(1-s)\cos[2(\theta_i-\theta_j)]$
(after \cite{GXY}). Interactions are purely nematic (resp. ferromagnetic) for $s=0$ (resp. 1). 
D, P, and N, respectively stand for disorder, polar order, and nematic order.} 
\label{fig1} 
\end{figure}

One might object that the properties of the ``nematic Vicsek model'' are {\it not} robust in the sense that the strict nematic symmetry of its interactions may be generically broken, albeit weakly, by ``friction'' effects during collisions between actual rods
(Fig.~\ref{fig1}a). 
On the other hand, in equilibrium, it is known that the quasi-ordered phase of the XY model with {\it nematic}
interactions resists a modest amount of ferromagnetic alignment 
(Fig.~\ref{fig1}b) \cite{GXY}. 

In this Rapid Communication, we introduce and study an out-of-equilibrium, Vicsek-style, version of such a
generalized XY model where ferromagnetic and nematic alignment compete.
We show that its phase diagram is similar to that of its equilibrium counterpart,
although rendered complicated by density-segregated sub-phases. Thus, the fully-nematic
Vicsek model is robust to the introduction of a modest bias of interactions towards ferromagnetic
alignment, alleviating the concern raised above for collisions of actual rods. We also show that
 this conclusion holds in the case of ``topological'' neighbors 
where interactions are not limited to some metric
zone but occur with those objects defining the first shell of Voronoi cells around the considered
particle. In both the metric and this ``metric-free'' case, we study the direct
polar-nematic transition present in phase diagrams like that of Fig.~\ref{fig1}b. We provide
evidence that it seems to be discontinuous in the metric model, but continuous in the 
metric-free case, with critical exponents of the Ising universality class, 
as in equilibrium \cite{GXY}. 
We finally discuss the inherent difficulties
in deriving hydrodynamic theories in the case of mixed ferromagnetic and nematic 
interactions.

Our starting point is a Vicsek-style model with competing ferromagnetic and nematic
interactions: $N$ point
particles move off-lattice at constant speed $v_0$ 
on a two dimensional $L \times L$ torus; particle $j$ is defined by its 
position $\mathbf{r}_j^t$ and 
orientation $\theta_j^t$, updated 
according to 
\begin{eqnarray} 
\label{motion_angle} 
\theta_j^{t+1}&=& 
\arg \left[\sum_{k\sim j} g_s(\theta_j^t,\theta_k^t)
\right] + \eta\,\xi_{j}^{t} \\ 
\mathbf{r}_j^{t+1}&=& \mathbf{r}_j^{t}+v_0\,  \mathbf{v}_j^{t+1}\, , 
\label{motion_pos} 
\end{eqnarray} 
where $\mathbf{v}_j^t=\left( \cos \theta_j^t, \sin \theta_j^t\right)^T$, 
the sum is taken over all particles $k$ within unit distance 
from particle $j$ (including $j$ itself),
$\xi$ is a white noise uniformly distributed in 
$\left[-\frac{\pi}{2},\frac{\pi}{2}\right]$, 
and the complex {\it stochastic} function $g_s(\theta,\theta')$ is:
\begin{equation} 
\label{eq:fp}
g_s(\theta, \theta')=\left\{
\begin{array}{ll}
e^{i\theta'}\;\;&{\rm with \;\; prob.}\;s\\
{\rm sign}\left[\cos(\theta -\theta') \right]e^{i\theta'}\;\;&{\rm  otherwise}
\end{array}\right.
\end{equation} 
Note that in the second case $g_s(\theta, \theta')$
 is invariant under  the transformation $\theta' \to \theta' + \pi$ and thus codes nematic alignment,
while the first case expresses ferromagnetic interaction.
For $s=0$, this model reduces to 
model studied in
\cite{Ginelli2010}, while $s=1$ is fully equivalent to the standard Vicsek model.
Thus, $s$ is a key parameter governing the relative weight of ferromagnetic interactions
which comes in addition to the two main parameters, the density $\rho=N/L^2$ and the noise strength
$\eta$. In the following, we focus on a low density system at $\rho=\frac{1}{8}$ with $v_{\rm 0}=\frac{1}{2}$, 
as in \cite{Ginelli2010}, and study the $(s,\eta)$ parameter plane. \cite{NOTE}

Systematic scans were performed for different sizes. While $L=512$ allows for a rough 
determination of the main features of the phase diagram, one needs larger sizes to capture
its details, mostly because the segregated banded states of the fully nematic ($s=0$) model 
arise clearly only for large systems \cite{Ginelli2010}.
Polar and nematic order were characterized by means of the two time-dependent 
global scalar order parameters $P(t)=|\langle\exp (i\theta^t_j)\rangle_j|$ 
(polar) and $Q(t)=|\langle\exp (i 2\theta^t_j)\rangle_j|$ (nematic), 
as well as their asymptotic time averages $P=\langle P(t)\rangle_t$ 
and $Q=\langle Q(t)\rangle_t$. To detect the various density-segregated phases, we relied
on snapshots and movies of coarse-grained density and order parameter fields.

\begin{figure}
\includegraphics[draft=false,clip=true,width=\columnwidth]{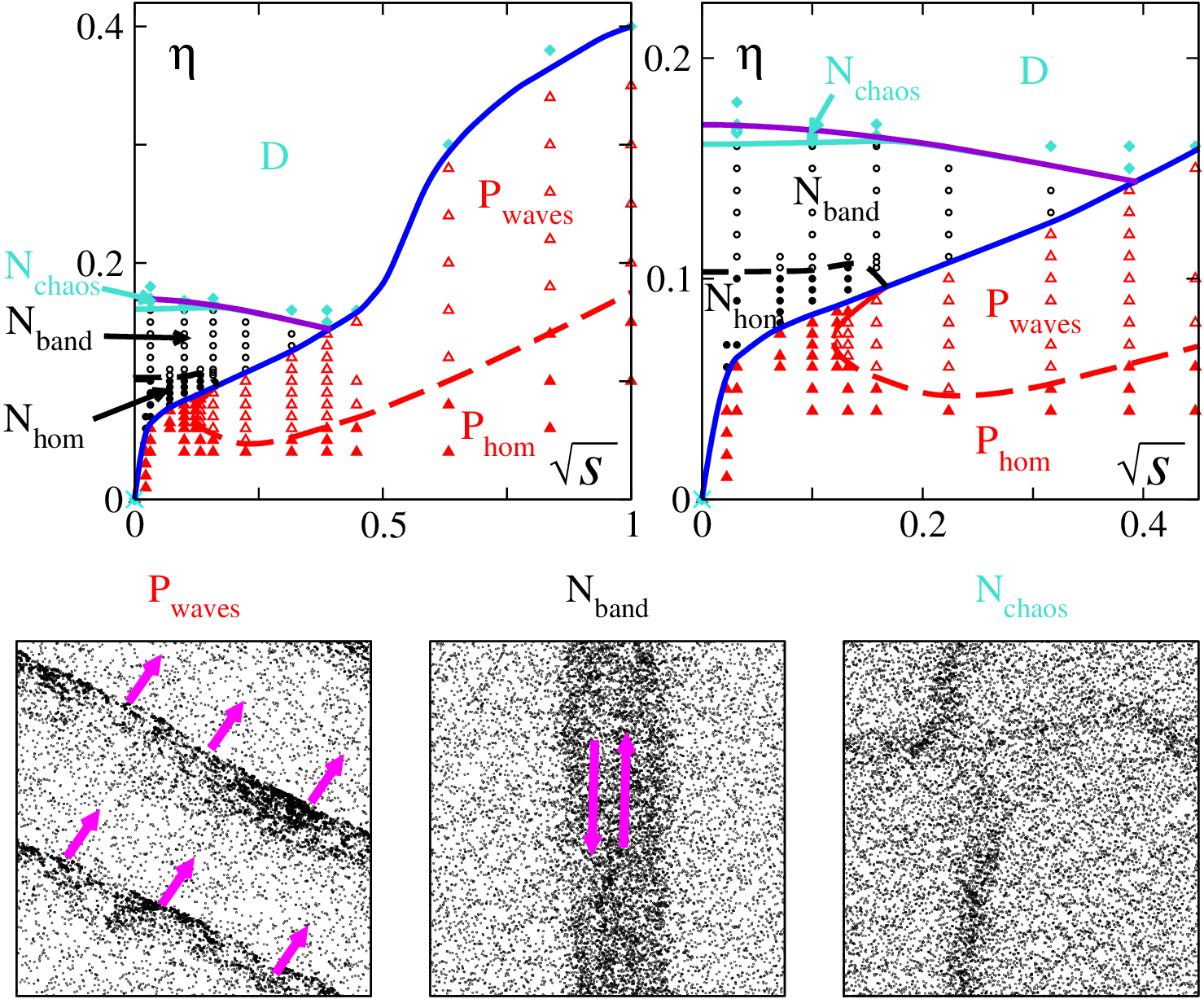}
\caption{(color online) Top panels: Phase diagram of the generalized Vicsek model with 
competing ferromagnetic and nematic interactions 
($\rho=\frac{1}{8}$, $L=1024$, $v_0=\frac{1}{2}$) 
(the top right panel is a close up of the left panel.). 
Results are presented in function of $\sqrt{s}$ for clarity. Symbols
located where individual runs were performed, coding the resulting observed phase:
red triangles for polar order, black circles for nematic order, and cyan diamonds for disorder. 
Full/empty symbols code for homogeneous/density segregated phases.
Bottom panels: typical snapshots of the 3 density segregated phases present (system size $L=1024$)
Left: ${\rm P}_{\rm waves}$ at $s=0.05$, $\eta=0.1$.
Middle: ${\rm N}_{\rm band}$ at $s=0.01$, $\eta=0.14$.
Right:  ${\rm N}_{\rm chaos}$ at $s=0.01$, $\eta=0.166$.
The thick magenta arrows indicate the direction of motion of particles.} 
\label{fig2} 
\end{figure}

The phase diagram is presented in Fig.~\ref{fig2} for $L=1024$, a size beyond which its various
features do not seem to change. 
Close to the $s=0$ axis, one observes that the various nematic sub-phases described in 
\cite{Ginelli2010} for the purely nematic case are extended to finite but small $s$ values: 
the spatially-homogeneous nematically-ordered phase (denoted ${\rm N}_{\rm hom}$) can be observed 
up to $\sqrt{s}\approx 0.17$ for $\eta\approx 0.1$. The segregated phase with a unique, dense, 
nematically-ordered band occupying a finite fraction of space (${\rm N}_{\rm band}$) extends up 
to $\sqrt{s}\approx0.4$ for $\eta\approx 0.13$.
The spectacular spatiotemporal chaos regime in which thin unstable dense bands elongate, twist, break,
collide and form again (denoted ${\rm N}_{\rm chaos}$) is limited to a narrow tongue near $\eta=0.16$.
As $s$ is increased, the width of this tongue decreases, making it difficult to locate numerically; our
results suggest, though, that it extends all the way to the meeting point with the polar order region.

Polar order can be observed at arbitrary small $s$, provided $\eta$ is small enough; 
it is delimited by a smooth line (like in the equilibrium case, see Fig.~\ref{fig2}) starting at the origin 
and ending at $\eta\simeq 0.4$ for $s=1$, 
the transition point of the Vicsek model at this density\cite{NOTE4}.
The polar order region is itself divided in two, as expected from our knowledge of the Vicsek model: 
in a large, tongue-like region bordering the onset of polar order  (${\rm P}_{\rm waves}$ in Fig.~\ref{fig2}), 
the ordered phase consists of trains of solitary traveling waves, and is thus different from
the  Toner-Tu homogeneous phase
(${\rm P}_{\rm hom}$) present at lower noise strength.

The phase diagram of our out-of-equilibrium, Vicsek-style, generalized XY model thus possesses 
the same general structure as its equilibrium counterpart: a small nematic triangular region 
is present on the small-$s$ side above a continuous line delimiting
polar order. This general structure is complicated by the presence of the various 
density-segregated ordered subphases. At our numerical resolution the line dividing the 
polar region (red dashed line in Fig.~\ref{fig2}) and that dividing the nematic
region (black dashed line) seem to meet the border of the polar region (blue solid line) 
at the same point \cite{NOTE2}. 
There are thus two different ways of transitioning directly from polar to nematic order, 
as opposed to only one at equilibrium: at small $s$ values, the P-N transition occurs between spatially-homogeneous
phases, while at intermediate $s$ values, it links the two density segregated phases 
${\rm P}_{\rm waves}$ and ${\rm N}_{\rm band}$ in Fig.~\ref{fig2}.

\begin{figure}
\includegraphics[draft=false,clip=true,width=\columnwidth]{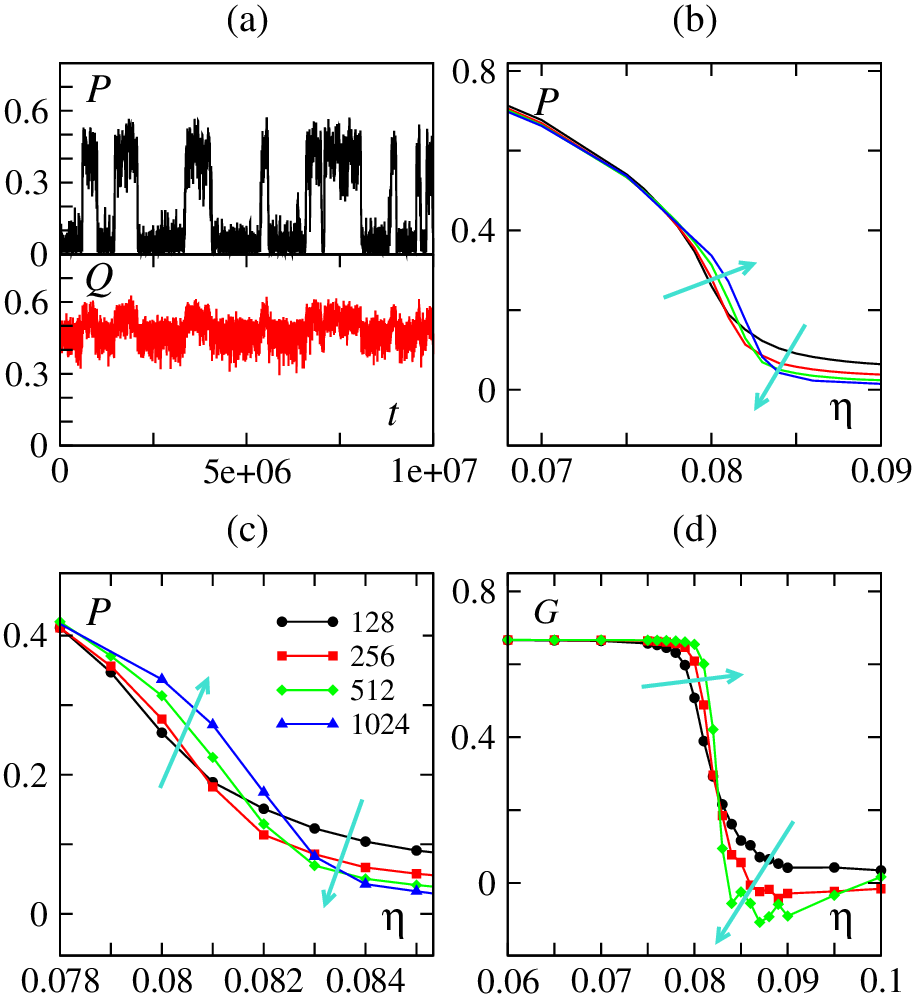}
\caption{(color online) Direct transitions from nematic to polar order in the generalized Vicsek model.
(a) transition between the segregated phases ${\rm P}_{\rm waves}$ and ${\rm N}_{\rm band}$ 
at $s=0.1$ occurring near $\eta=0.132$ (time series of polar (top) and nematic (bottom) 
order parameter for $L=256$).
(b-d) transition between the homogenous phases ${\rm P}_{\rm hom}$ and ${\rm N}_{\rm hom}$ 
at $s=0.01$ occurring near $\eta=0.085$.
(b,c): order parameter curves at sizes $L=128$, 256, 512, 1024. Panel (c) is a close-up of panel (b).
(d): Binder cumulant curves at the same system sizes. The arrows indicate increasing system size.} 
\label{fig3} 
\end{figure}

We now turn our attention to the nature of the main transitions 
(disorder/nematic, disorder/polar, and polar/nematic). 
Let us first recall that the ordered phases observed all seem numerically
to show true long-range order, even the nematic phases  ${\rm N}_{\rm hom}$ and  ${\rm N}_{\rm band}$
\cite{Ginelli2010}. In these last cases, however, as argued in \cite{Ginelli2010}, 
one cannot exclude the possibility that nematic
order is only quasi-long-range at very large scales. In the following, we assume true long-range order,
in agreement with the numerical results.

The transition from disorder to nematic order actually occurs between the chaotic 
${\rm N}_{\rm chaos}$ phase and the ordered, segregated 
${\rm N}_{\rm band}$ phase. As reported in \cite{Ginelli2010}, 
it is determined by the long-wavelength instability of the band, which leads to the 
${\rm N}_{\rm chaos}$ phase. This phase being disordered, albeit with very large intrinsic scales,
the nematic order parameter $Q$ shows a discontinuous jump at the transition.
The disorder/polar transition occurs between the microscopically disordered phase D 
and the segregated phase ${\rm P}_{\rm waves}$. Like for the original Vicsek model, 
it is also discontinuous, as the polar order parameter takes finite, order 1 values as soon
as the waves appear \cite{Chate2008}.

We studied the two different P-N transitions present. 
The transition between the segregated
phases ${\rm P}_{\rm waves}$ and ${\rm N}_{\rm bands}$ was studied as a function of 
$\eta$ at $s=0.1$. It is clearly discontinuous, as seen, e.g., 
in the characteristic flip-flop dynamics of the polar order parameter $P$ in the 
transition region, leading to a bimodal distribution testifying of phase coexistence
 (Fig.~\ref{fig3}a).
 The transition between the homogeneous nematic and polar phases is more difficult to characterize. 
We performed a finite-size scaling study at $s=0.01$, varying $\eta$ 
around the transition point.
The behavior of the $P(\eta)$ 
curves (Fig.~\ref{fig3}b) indicates the premises of a discontinuous transition: 
although no discontinuity proper is present even at the largest
size considered, these curves cross each other, suggesting that a jump may appear 
at still larger sizes. This conclusion is also borne out of the behavior of the 
Binder cumulant $G=1-\langle P(t)^4 \rangle_t/(3\langle P(t)^2 \rangle^2_t )$, 
which develops a deeper minimum as the system size 
is increased (Fig.~\ref{fig3}d) \cite{FSS}.

We now report on the properties of the metric-free, ``topological" version of our 
generalized Vicsek model with competing ferromagnetic and nematic alignment. This case is of theoretical
interest because no density-segregated phases are present in the purely ferromagnetic or nematic cases
\cite{TOPOVICSEK,TOPOKINETIC,MCM-Gopinath}, 
opening the way to continuous phase transitions (see below).
It is also relevant in the context of
collective motion. Common sense and some experimental/observational
evidence indicate that in groups of higher organisms moving collectively 
(bird flocks, fish schools, crowds, etc.), one individual, rather than interacting with neighbors 
located within a given metric zone around itself, takes into account those individuals forming 
some ``angular landscape" to perform navigation decisions \cite{Moussaid2011}. 
It was suggested for instance that starlings interact with their 7-8 nearest neighbors, 
irrespective of the flock density \cite{Starlings}. 
Careful study of fish trajectory data showed that in some schools the stimulus/response function
of fish can be modeled by interactions with their first Voronoi neighbors
(those whose polygons, in a Voronoi tessellation of space, form the first shell 
around the focal fish) \cite{Gautrais2012}.
Here, following a study of a metric-free version of the Vicsek model \cite{TOPOVICSEK}, 
we consider these Voronoi neighbors, and we use ``vectorial'' noise, which means that 
Eq.(\ref{motion_angle}) is replaced by
$\theta_j^{t+1}= 
\arg \left[\sum_{k\sim j} g_s(\theta_j^t,\theta_k^t)+ \eta\,{\cal N}_j\vec\xi_{j}^{t}\right]$
where $\vec\xi$ is a randomly oriented unit vector and ${\cal N}_j$ is the current number of (Voronoi)
neighbors of particle $j$.

\begin{figure}
\includegraphics[draft=false,clip=true,width=\columnwidth]{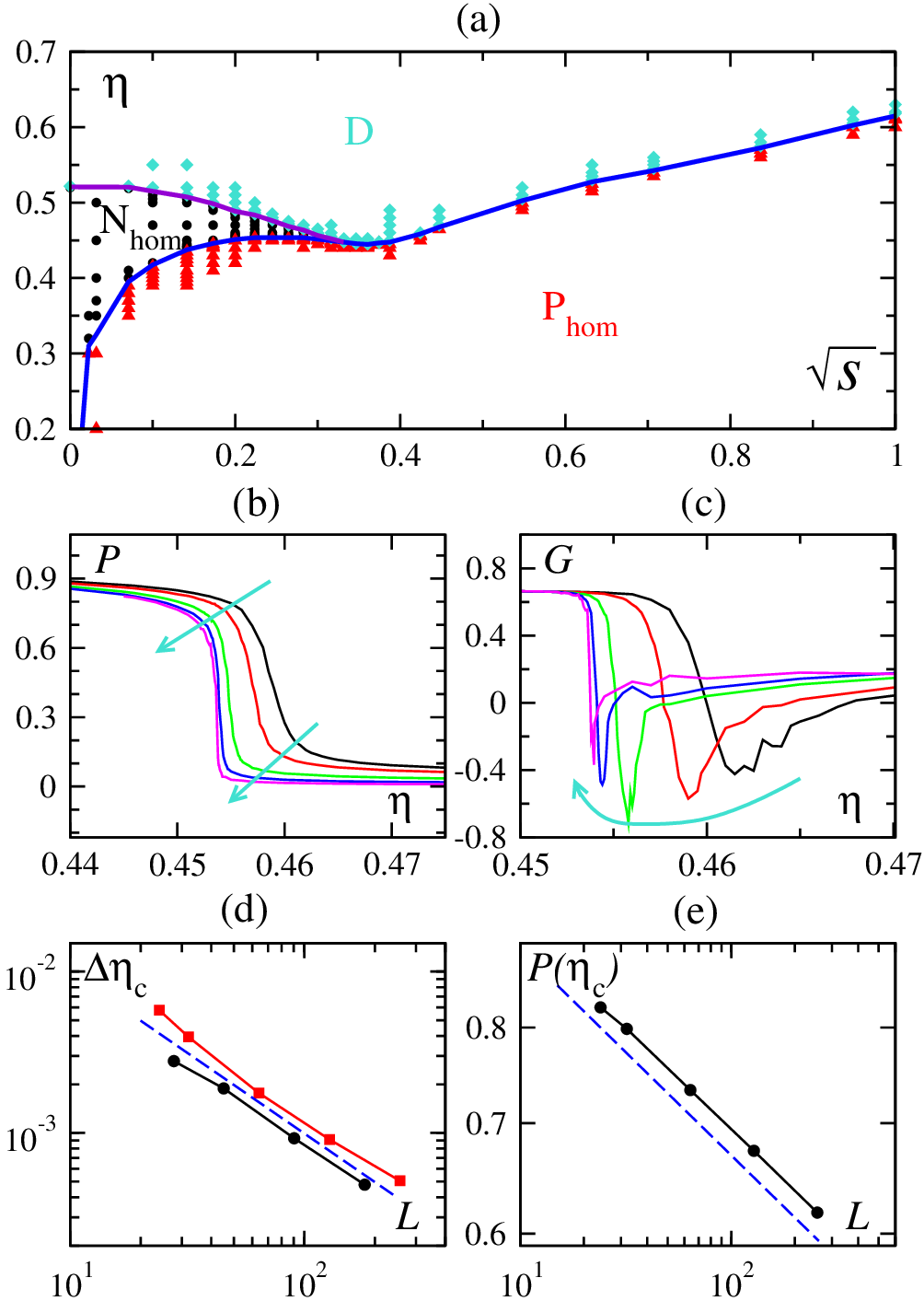}
\caption{(color online) Metric-free version of the model where particles interact 
with their Voronoi neighbors.
(a): phase diagram obtained for $L=128$. Legends as in Fig.~\ref{fig2}ab.
(b-d): finite-size scaling study of the P-N transition occuring at $s=0.05$.
(b) and (c): $P(\eta)$ and $G(\eta)$ for $L=24$, 32, 64, 128, and 256 
(the arrows indicate increasing system size). The order parameter curves do {\it not} 
cross each other; the Binder cumulant curves show a minimum which first deepens then start
receding as $L$ increases. These curves do cross each other when $G\sim\frac{2}{3}$ (not shown).
(d): scaling of finite-size distance to asymptotic threshold 
$\Delta\eta_{\rm c}(L)=|\eta_c(L)-\eta_c^\infty|$
for estimated threshold $\eta_c^\infty=0.4530(1)$. Red squares: $\eta_{\rm c}(L)$ is the 
location fo the maximum of susceptibility at size $L$. Black circles: $\eta_{\rm c}(L)$ is the 
crossing point of two $G(\eta)$ curves at sizes $L_1$ and $L_2$ with $L=\sqrt{L_1L_2}$.
The dashed blue line has slope -1.
(e): scaling of order parameter at $\eta=\eta_{\rm c}^\infty$. The dashed blue line has slope 
$-\frac{1}{8}$.} 
\label{fig4} 
\end{figure}

As shown in \cite{TOPOVICSEK,TOPOKINETIC}, 
global density drops out of the problem in metric-free models (it can be scaled out). 
In practice we worked at unit density, with $v_0=0.5$. 
The phase diagram of this metric-free version of our generalized Vicsek model is shown 
in Fig.~\ref{fig4}a for $L=128$ (diagrams obtained at larger sizes are nearly indistinguishable).  
As expected, no density-segregated phases are present, 
leaving a diagram qualitatively similar to that of the equilibrium case. The P-D and N-D transitions are now continuous, as expected from \cite{TOPOVICSEK,TOPOKINETIC} where the $s=1$ and $s=0$ cases
were studied. A detailed study of the associated critical exponents will be presented elsewhere. 
(For the P-D transition, they are consistent with the values reported in \cite{TOPOVICSEK}.)
We studied the direct P-N transition at $s=0.05$ by finite-size scaling. 
The (polar) order parameter curves now do {\it not} cross each other (Fig.~\ref{fig4}b), 
and the Binder cumulant curves show a minimum which, after deepening at small sizes, eventually start
receding at the largest sizes we could probe (Fig.~\ref{fig4}c). 
These qualitative facts point to a continuous transition. Quantitatively, our estimates of critical
exponents are as follows:
The crossings of the Binder cumulant curves (near $G\sim\frac{2}{3}$, not visible on Fig.~\ref{fig4}c)
converge to an asymptotic threshold $\eta_{\rm c}=0.4530(1)$
with an exponent $1/\nu=1.00(5)$. 
The location of the maxima of the susceptibility also converge to the same estimated threshold
with the same estimated $1/\nu$, in excellent agreement with the Ising
value $\nu=1$ (Fig.~\ref{fig4}d). The peak values of the susceptibility diverge with system
size with exponent $\gamma/\nu=1.74(2)$, the Ising value being $\frac{7}{4}$ (not shown). 
The order parameter decreases algebraically at the estimated critical point 
with exponent $\beta/\nu=0.126(3)$, 
in close agreement with the Ising value $\frac{1}{8}$ (Fig.~\ref{fig4}e) .
Thus, in spite of rather strong finite-size effects as testified by the behavior of the
Binder cumulant curves on the nematic side, all estimated critical exponents are very close to
their Ising universality class values, as expected from studies of the equilibrium 
generalized XY model \cite{GXY}. 

Before summarizing, we discuss briefly the derivation of a continuous theory 
describing active matter systems with competing
ferromagnetic and nematic alignment interactions. Such a theory must a priori be in terms of
a polarity field ${\bf P}$ and a nematic tensorial field  ${\bf Q}$ if it is to account for 
nematically-ordered phases. Baskaran and Marchetti have proposed rather complicated such equations for
the case of self-propelled rods interacting via steric exclusion \cite{MCM-NEMA}. We have followed
the somewhat simpler route of the ``Boltzmann'' approach used in \cite{BDG,TOPOKINETIC}, which is 
particularly adapted to dilute Vicsek-like models, for the case of competing ferromagnetic and nematic
alignment\cite{NOTE3}. While details will be published elsewhere \cite{TBP},
we only mention here that the obtained equations for  ${\bf P}$ and  ${\bf Q}$ fail to account
for the phase diagrams presented here. They confirm the existence of a nematic phase at finite $s$, but
are not suitable to describe the polar phase, probably since the truncation scheme used is only valid 
near onset of nematic order.

In conclusion, we have studied a Vicsek-style model with competing ferromagnetic and nematic alignment 
in both a metric and a ``metric-free'' version where interactions take place with Voronoi
neighbors. We have shown that the fully nematic case of this out-of-equilibrium XY model 
(``self-propelled rods'') resists some bias toward ferromagnetic alignment, thus conferring some
robustness to its nematically-ordered phases and allaying our initial concern about friction effects
inducing a ``polar bias'' in aligning collisions of elongated objects in experiments.
In the metric case, the direct polar/nematic transition
has been found discontinuous, in line with the order/disorder transitions,
whereas the metric-free version exhibits an Ising-class continuous transition, 
as in equilibrium. We have signalled that a simple derivation of
a continuous theory able to account for all observed facts is not easy, and constitutes an important
future step for putting these results on firmer ground.

\acknowledgements
We thank Eric Bertin and Anton Peshkov for useful discussions. This work was largely performed
within the activities of the Advanced Study Group ``Statistical Physics of Collective Motion'' at 
the Max Plank Institute for the Physics of Complex Systems, Dresden, Germany.
FG acknowledges support by Grant IIT-Seed Artswarm.


\begin{thebibliography}{99}
\bibitem{NOTE} We do not expect significant changes for different densities, but a full exploration
of parameter space remains to be done.
\bibitem{NOTE4} There is a factor 2 between the noise strength definition used here and that used in
\cite{Chate2008}, where this threshold was found near 0.2.
\end{thebibliography}
\end{document}